\documentclass[aps,prl,twocolumn,amsmath,amssymb,nofootinbib,superscriptaddress,floatfix]{revtex4}
\usepackage{graphicx}
\usepackage{amssymb,amsmath,bbm,braket,indentfirst,bm,hyperref}
\usepackage{epstopdf}
\usepackage{dcolumn}
\usepackage{color}
\usepackage{bigstrut}
\usepackage{empheq}
\usepackage{tcolorbox}
\usepackage{mathtools}

\def \>{\rangle} 
\def \<{\langle} 
 
\newcommand{\be}{\begin{equation}}
\newcommand{\ee}{\end{equation}}
\newcommand{\bea}{\begin{eqnarray}}
\newcommand{\eea}{\end{eqnarray}}

\begin{document}

\title{Data-driven modeling reveals a universal dynamic underlying the COVID-19 pandemic under social distancing}
\author{Robert Marsland III}
\email{marsland@bu.edu}
\affiliation{Department of Physics, Boston University, Boston, Massachusetts 02215, USA}
\affiliation{Biological Design Center, Boston University, Boston, Massachusetts 02215, USA}
\affiliation{Microbiome Initiative, Boston University, Boston, Massachusetts 02215, USA}

\author{Pankaj Mehta}
\email{pankajm@bu.edu}
\affiliation{Department of Physics, Boston University, Boston, Massachusetts 02215, USA}
\affiliation{Biological Design Center, Boston University, Boston, Massachusetts 02215, USA}
\affiliation{Microbiome Initiative, Boston University, Boston, Massachusetts 02215, USA}
\affiliation{Bioinformatics Program, Boston University, Boston, Massachusetts 02215, USA}
\affiliation{College of Data Science, Boston University, Boston, Massachusetts 02215, USA}

\date{\today}

% FORMAT NOTE: Nature Physics Abstract limit 150. PRL is 600 characters including spaces
% CURRENT STATS: 81 words, 611 characters with spaces
\begin{abstract}
We show that the COVID-19 pandemic under social distancing exhibits universal dynamics. The cumulative numbers of both infections and deaths quickly cross over from exponential growth at early times to a longer period of power law growth, before eventually slowing. In agreement with a recent statistical forecasting model by the IHME, we show that this dynamics is well described by the erf function. Using this functional form, we perform a data collapse across countries and US states with very different population characteristics and social distancing policies, confirming the universal behavior of the COVID-19 outbreak. We show that the predictive power of statistical models is limited until a few days before curves flatten, forecast deaths and infections assuming current policies continue and compare our predictions to the IHME models. We present simulations showing this universal dynamics is consistent with disease transmission on scale-free networks and random networks with non-Markovian transmission dynamics.

\end{abstract}
\maketitle

\section{Introduction}

The COVID-19 pandemic represents a unique challenge for understanding and predicting the dynamics of disease spreading and fatalities. The large number of individuals infected, the global nature of disease transmission, and the limited ability to test individuals highlight the importance of formulating effective mathematical and statistical models for forecasting disease dynamics.  Fortunately, the challenges posed by COVID-19 have also led to an impressive effort at collecting and centralizing data on infections and fatalities across countries and regions \cite{dong2020interactive}. Using this data, there have been numerous efforts to understand and forecast the disease dynamics of the COVID-19 pandemic.

These efforts fall into two major categories. The first approach is based on fitting this data to classical models from mathematical epidemiology such as SIR models, with various level of detail and complexity  \cite{hethcote2000mathematics, diekmann2012mathematical}. A major advantage of this approach is that the dynamics of SIR models can be parameterized using only a few parameters, most notably $R_0$, the average number of individuals that will be infected by an individual who has the disease \cite{lipsitch2003transmission,wu2020estimating}. This allows for concrete predictions \cite{maslov2020window} and assessment of how mechanistic processes such as seasonal variation \cite{neher2020impact, kissler2020projecting} and social policies \cite{wallinga2010optimizing} affect disease dynamics. However, such mathematical models approaches also suffer from a number of drawbacks. Many of these models often assume well-mixed populations (with or without population structure), implicitly ignoring the effect of network and spatial structure \cite{newman2002spread, barthelemy2005dynamical} . While many of these difficulties can be overcome by considering more refined models \cite{pastor2015epidemic, balcan2011phase, chinazzi2020effect}, this requires access to detailed data that is often unavailable.

A second widely used approach for modeling the COVID-19 pandemic is to use statistical models to make predictions. A particularly prominent example of this are recent forecasts by the  Institute for Health Metrics and Evaluation (IHME) at the University of Washington, made by fitting the number of deaths as a function of time using the cumulative distribution function of the normal distribution (we will use the shorthand ``erf function'' for this functional form through out the manuscript) \cite{covid2020forecasting}. This approach has yielded tangible and sensible predictions and has been used to guide policy by state and national governments. However, unlike the mathematical models discussed above, it remains unclear why and when the erf function gives a good fit to disease dynamics. In particular, there is no mechanistic understanding of what epidemiological processes give rise to this fitting form or how to incorporate the effect of various interventions on disease propagation. Understanding the answers to these pressing questions is especially important given the significant role that these models are playing in guiding public policy.

To begin addressing these shortcomings, in this paper we make use of the extensive data on COVID-19 infections and death across regions, US states, and countries to show that COVID-19 spread follows a universal dynamics: exponential growth at early times (less than $\sim 10$ days after the fifth death), followed by a longer period where infections and deaths grow as a power law, before eventually saturating at later times. We show that this entire dynamics is well fit by the erf function, allowing us to perform a data collapse on dynamics across countries and US states. Surprisingly, this single functional form can capture the dynamics of disease progression in regions/countries with extremely different population characteristics and social policies. Nevertheless, we show that it is difficult to make accurate predictions until a few days before infections peak. We compare and contrast our predictions to the IHME model and present plots integrating our predictions with the country-specific  social stringency index compiled by Oxford COVID-19 Government Response Tracker \cite{OxfordCOVID19}.   We show that the erf model follows naturally from considering transmission on social networks. We conclude by discussing the practical implications of our results. 

\section{Assumptions and Limitations}
Before presenting our results, we briefly review the assumptions and limitations of our approach. 
\begin{itemize}

\item We stress first of all that the present work only deals with disease dynamics under the policies of social distancing currently implemented in each country. Understanding this existing data is an essential prerequisite for adequately answering the urgent practical question about the expected results of relaxing current policies. The main contribution of the present analysis is to provide a unified framework for summarizing the characteristics of these dynamics in each country and showing that each timeseries can be represented by just two parameters. This provides a foundation for further analysis of how social policies, population characteristics and other factors affect the values of these parameters. 

\item Secondly, we emphasize that the confidence intervals we provided on our predictions are lower bounds on the true amount of uncertainty. These intervals are calculated under the assumption that the data is accurately described by the fitting function given in Eq. (\ref{eq:fittingfunction}) below, with no changes in parameter values, and with all deviations from this function due to random multiplicative noise. The confidence intervals do not account for possible future changes in social policy, or for systematic deviations from the fitting function that may arise at late times. In particular, unlike prominent models such as those from the IHME, \emph{ we make no attempt to model the effects of social policy on infections and deaths}. Instead, we seek to understand the differences and similarities of disease spreading dynamics across countries given their current social policies. For this reason, these models are unable to answer questions about what would happen if a country adopted social policies that were dramatically different from those that are currently being pursued in our dataset.

\item Finally, we point out that variability in reporting standards and other issues of reliability necessarily affect any real-time global data aggregation effort. This is especially true of the test result data which depends on the availability of testing in each region as well as on government policies for administration of the tests. For this reason, we focus on fatality data in the main text, which is commonly taken to be more reliable. Nevertheless, the standards for attributing a fatality to COVID-19 vary significantly from region to region and over time, complicating the interpretation of these data sets as well. Despite this variation,  the observed universal curves for both fatalities and cases indicate that meaningful information on the intrinsic disease dynamics can still be extracted from these reports for many countries.

\end{itemize}

\begin{figure*}[ht]
    \centering
    \includegraphics[width=15cm]{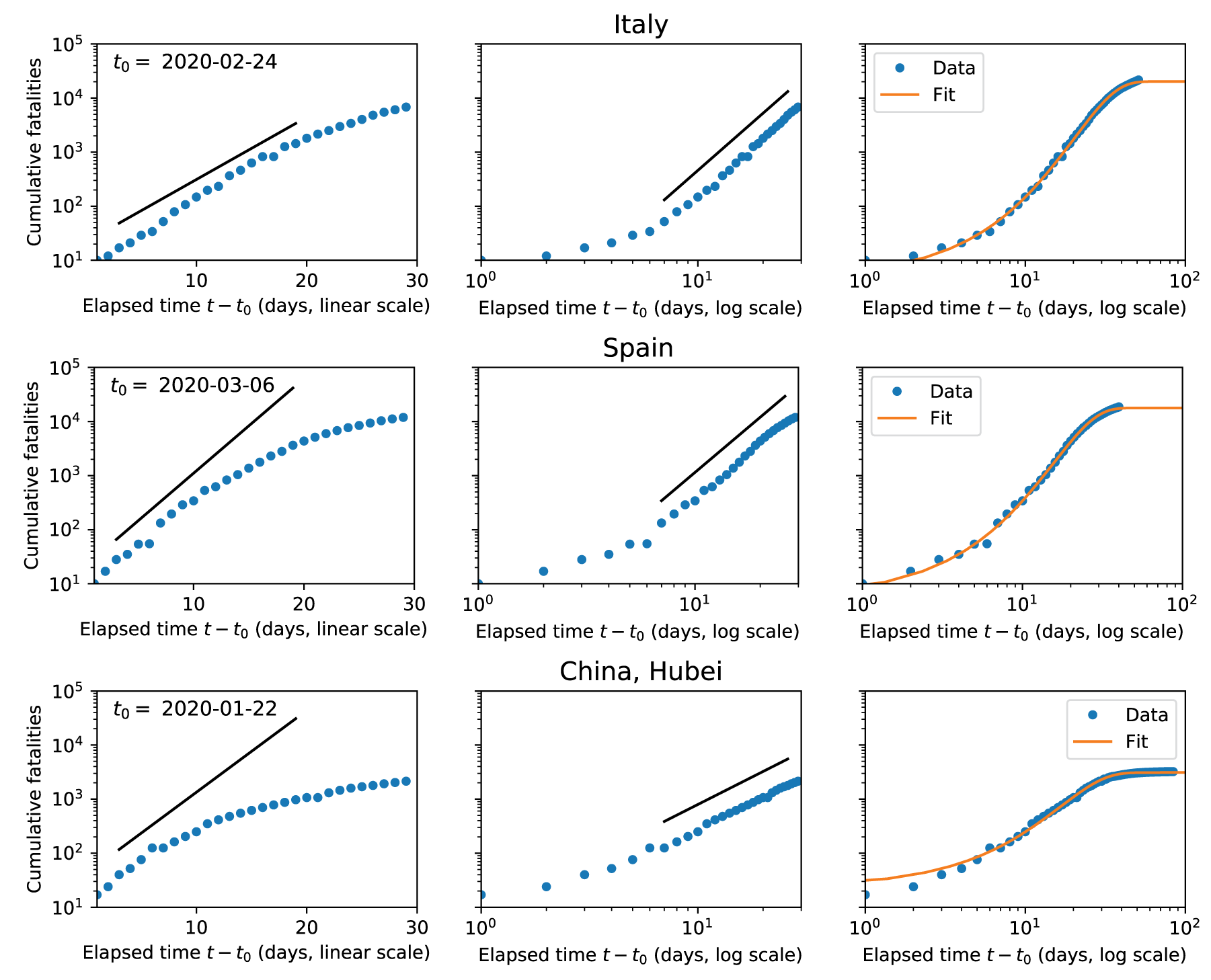}
    \caption{{\bf Post-peak countries show power-law dynamics at intermediate times}. Cumulative fatalities (blue points) versus time for three countries that passed their peak rate of new deaths before April 1, 2020. First two columns show the dynamics for 30 days after the day $t_0$ when the number of cumulative fatalities first exceeded 5. The time axis in the first column is on a linear scale, while in the second it is logarithmic. Fatalities are shown on a logarithmic scale for both columns. Straight lines are included for reference, indicating an initial phase of exponential growth (straight line on log-linear plot) followed by a phase of power-law growth (straight line on log-log plot). Third column shows cumulative normal distribution, Eq. (\ref{eq:fittingfunction}), fit to the entire timeseries (through April 15, 2020), on log-log axes. The day $t_0$ when the number of fatalities first exceeded 5 is indicated in each row.}
    \label{fig:CIS}
\end{figure*}

\section{Results}

We analyzed the US and global time series data on infections and deaths from the Johns Hopkins University Coronavirus Resource Center, which are updated daily \cite{dong2020interactive}. Since we were interested in universal dynamics, we only used times after the cumulative number of confirmed cases in the region exceeded 500 for infection data, or the cumulative number of fatalities exceeded 50 for death data (see Materials and Methods). We use these cutoffs to minimize the effects of stochastic fluctuations arising from small numbers of infected individuals at early times. We required at least 6 data points to remain in the time series after truncation in order to proceed with model fitting. This limited our analysis to 79 out of 322 total regions and countries in the global data, including 30 out of fifty US states. Since the data is updated daily, in all our analysis we limit ourselves to data up to April 15 (though the scripts can be rerun easily for the most current data set). All code is available as Juypter Notebooks on Github at \url{https://github.com/Emergent-Behaviors-in-Biology/covid19}. In the main text, we focus on plots of cumulative deaths. Analogous plots for cumulative cases can be found in the appendix.  

\begin{figure*}[ht]
    \centering
    \includegraphics[width=15cm]{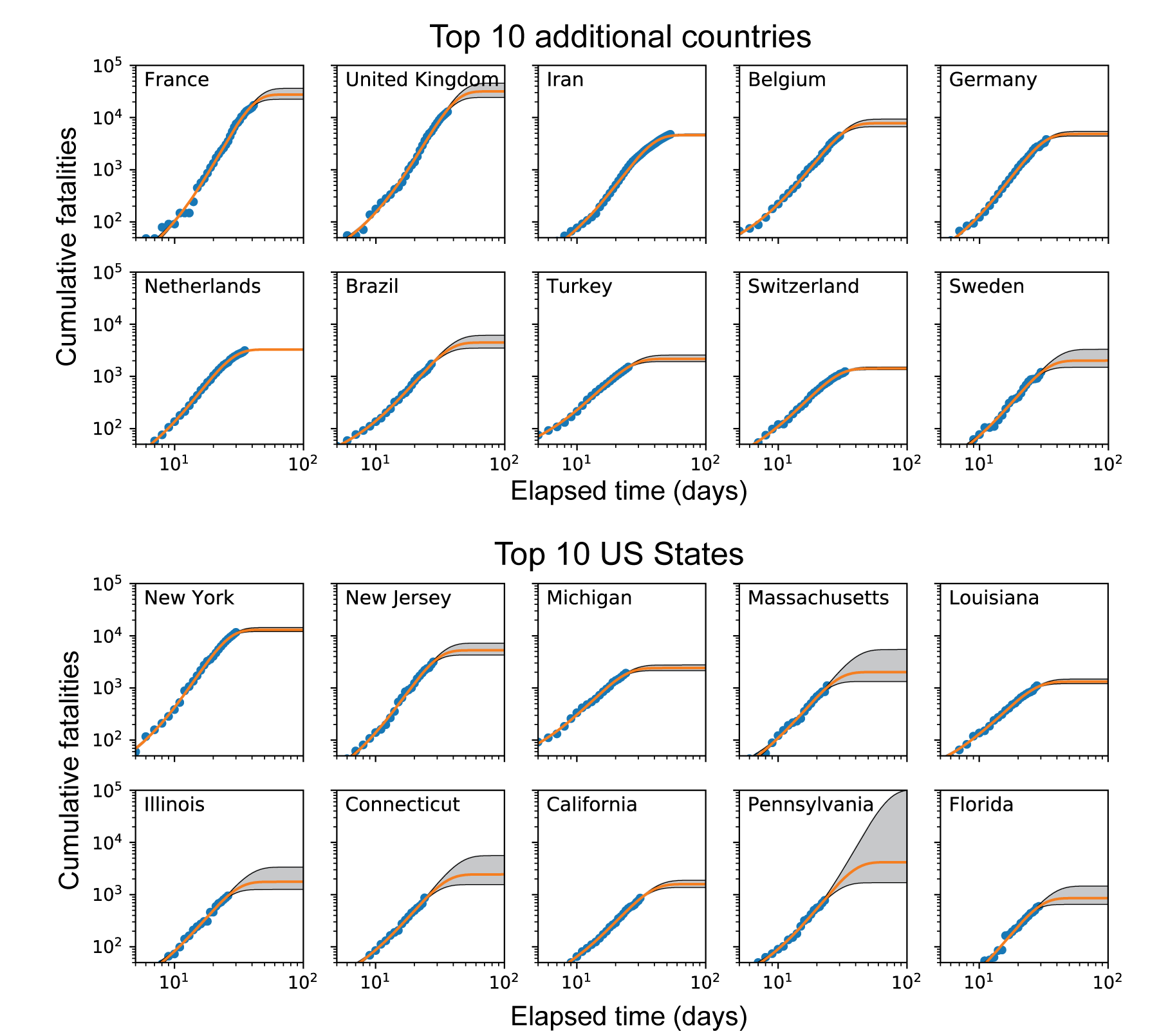}
    \caption{{\bf Power law behavior observed across many different countries.} Log-log plots of cumulative fatalities for the ten countries (excluding Italy, Spain and China, which were shown above) and ten states with most fatalities. Orange line is best fit to the cumulative normal (Eq. \ref{eq:fittingfunction}), and gray region is the confidence interval computed as described in the Methods. Note that this confidence interval is a lower bound on the true amount of uncertainty, since the computation assumes that the data is accurately described by the fitting function given in Eq. (\ref{eq:fittingfunction}) below, with all deviations from this function due to random multiplicative noise. The uncertainty estimation does not account for possible future changes in social policy, or for systematic deviations from the fitting function that may arise at late times.}
    \label{fig:predall}
\end{figure*}

\subsection{Power law growth dominates early COVID-19 disease dynamics}

We begin by plotting the cumulative cases and deaths of three regions/countries whose death rate curves had flattened by Wednesday, April 1: Hubei, China, Italy, and Spain. Many models and discussions of disease dynamics emphasize the exponential growth of cases at early times. If cumulative deaths and cases are growing exponentially, then the data should fall on a straight line in a semi-log plot (i.e. log number of deaths vs time). The left hand panels of Figure \ref{fig:CIS} show semi-log plots of cumulative deaths as a function of time for China, Italy, and Spain (see left panel of Fig. \ref{fig:sup1} for analogous panels for the number of confirmed cases). As expected, at early times the dynamics follow an exponential growth curve and are well fit by a straight line on the log-log plot. However, after about ten days, the data quickly deviates from the straight line, indicating a slowdown in the growth of the number of deaths and cases.

The data for all three countries quickly transitions to power law growth, where the cumulative number of deaths (or confirmed cases), $N$, grows as a power law with time, $t$. In other words, we have that $N \propto t^\alpha$, with larger $\alpha$ meaning faster growth. Power-law growth is easiest to see in a log-log plot since $\log N \propto \alpha \log t$, meaning that the data should fall on a straight line with slope $\alpha$ on a log-log plot.  The middle panels of  Figures  \ref{fig:CIS} and \ref{fig:sup1} show the cumulative deaths and confirmed cases on a log-log plot. It is clear that for about $20$ days the data is well described by power-law growth before eventually flattening. This is consistent with previous observations of early power-law dynamics in other epidemics including HIV/AIDS and the 2014 Ebola outbreak \cite{viboud2016generalized,chowell2016mathematical}.

\subsection{Data is well fit by the erf function}

\begin{figure*}[t]
    \centering
    \includegraphics[width=15cm]{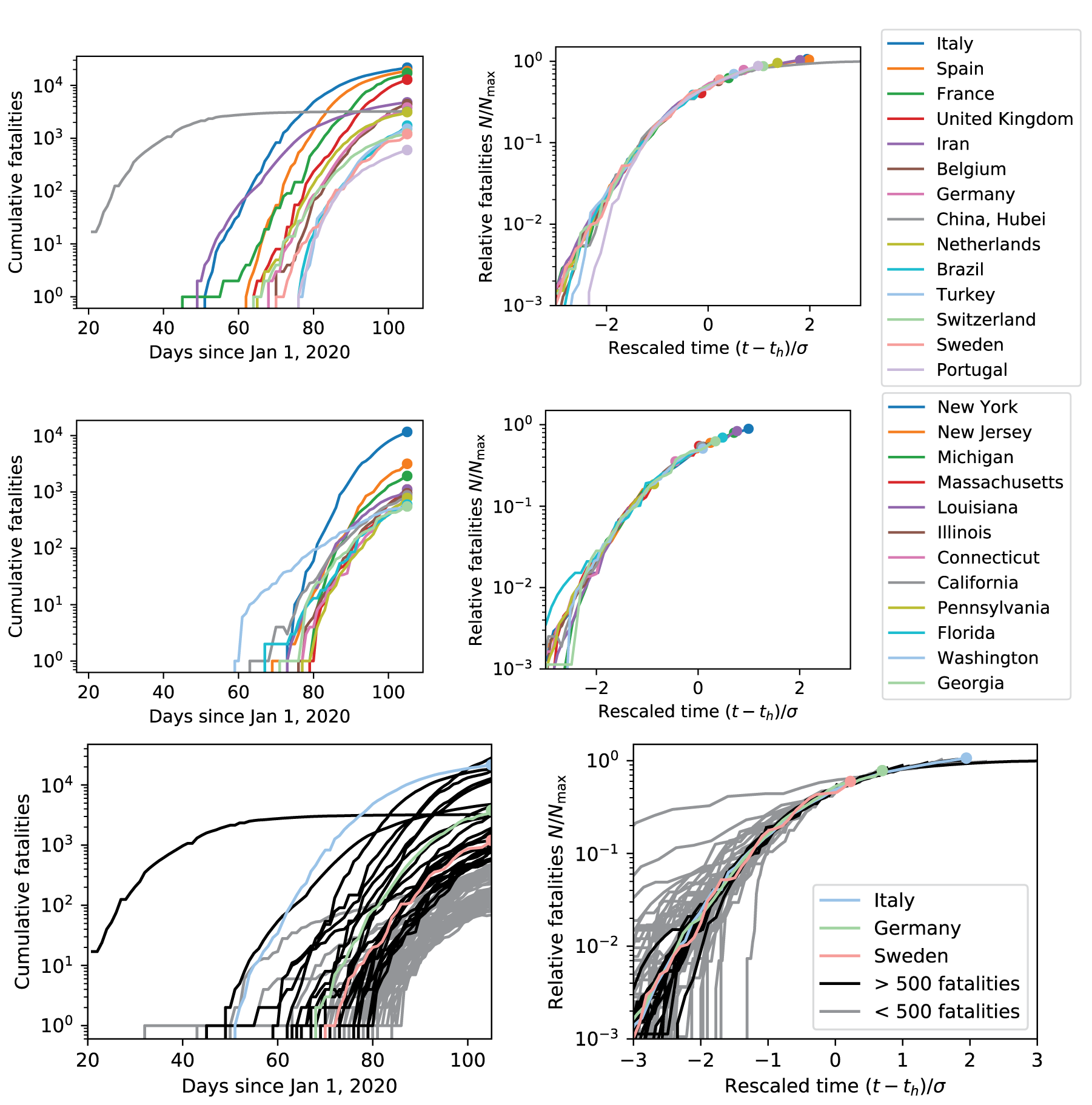}
    \caption{{\bf Rescaling time and fatalities using fit parameters yields universal curve}. Top: Raw (left) and rescaled (right) cumulative fatality data for all countries except for US with more than 500 fatalities as of April 15, 2020. Rescaled vertical axis (``Relative fatalities'') is the cumulative number of fatalities $N$ divided by the best-fit value of $N_{\rm max}$ for each country. Rescaled time axis is the time relative to the best-fit half-max time $t_h$, divided by the timescale $\sigma$. Middle: Same as above, but for US states with more than 500 fatalities as of April 15, 2020. Bottom: Same as above, but for all regions (countries and states/provinces) with sufficient data to fit.}
    \label{fig:collapse}
\end{figure*}

While the exponential followed by power law growth can capture the early growth of cases, these functions are incapable of capturing the full disease dynamics -- in particular, the eventual flattening and saturation of the number of confirmed deaths and cases. Recently,  researchers from the IHME noted that the disease dynamics is well fit by the erf function  (more precisely the cumulative distribution function of the normal distribution) \cite{covid2020forecasting}. Surprisingly, the IHME found that the erf function yields much better fits to the data than a sigmoid (another saturating function) even though the sigmoid function is thought to be the natural ``mean-field'' description of disease propagation of homogenous social interaction networks \cite{anderson1992infectious, murray2007mathematical, barthelemy2005dynamical}. Our initial data exploration confirmed this observation (data not shown) and for this reason we focused on fitting our full curves using the erf function suggested in \cite{covid2020forecasting}.

This fitting function is a saturating function defined by three  parameters: the final number of fatalities/cases $N_{\rm max}$ that will occur in the region, the time $t_h$ at which cumulative confirmed fatalities/cases reach $N_{\rm max}/2$, and a parameter $\sigma$ that controls the natural time scale of the infection dynamics. Note that $t_h$ is also the time at which the rate of new fatalities/cases reaches its peak. In general, these parameters must be inferred directly from the data in each region or country one wants to make predictions about. Once the parameters are inferred, the disease dynamics under current social and policy conditions are fully specified, allowing for forecasting predictions \emph{under the assumption that conditions remain the same and there are no dramatic changes in social policy}. Uncertainty in the values of the parameters naturally translates to uncertainty in predictions since small changes in these parameters can often result in very different final dynamics. A detailed description to the fitting procedure can be found in the Materials and Methods. In brief, the data was fit 
to the functional form
\be
N(t) = N_{\rm max}\left[\frac{1}{2}+{\rm erf}\left(\frac{t-t_h}{\sqrt{2}\sigma}   \right) \right],
\label{eq:fittingfunction}
\ee
where
\be
\mathrm{erf}(t) ={1 \over \sqrt{\pi}} \int_0^t e^{-x^2}\,dx,
\ee

Our fitting procedure differs from recent forecasts by the IHME in several crucial ways \cite{covid2020forecasting}. First, the IHME used a Generalized Linear Model with linking functions on the whole global dataset, while we directly fit each region independently. We make this choice because of the very different reporting strategies and criteria used by different governments and regions. Secondly, the IHME incorporates  prior expectations about the disease dynamics into their model, by assuming (1) that the infection time scale $\sigma$ of all countries and regions should be similar to that of Wuhan, China, and (2) that either the total number of fatalities $N_{\rm max}$ or the interval $t_h-t_0$ between the beginning of the infection and the peak time depend on the timing of social distancing. In contrast, our model makes no assumptions about the infection time scale and is agnostic about the effects of social distancing.  It assumes as a given the spectrum of social policies adopted by various regions and countries whose time series we fit. Finally, to estimate error bars we make use of the fact that the Fisher Information (inverse Hessian of the cost function) for our cost functions are effectively one-dimensional (see Material and Methods and Figure \ref{fig:params}). This simplifies our calculations of confidence intervals.

The best fits of Eq. (\ref{eq:fittingfunction}) to the data from Italy, Spain and Hubei are shown in the orange lines in the right hand panels of Figures  \ref{fig:CIS} and \ref{fig:sup1}. As can be seen from the plots, this functional form yields incredibly good fits to all time series. However, any realistic forecasting model also requires one to estimate uncertainty about the prediction parameters. For this reason, it is extremely important to also provide uncertainty estimates when fitting the data. Figure \ref{fig:predall} shows fits to the erf function with corresponding uncertainty intervals for countries with over 500 fatalities and the ten US states with the most deaths. Notice, that while the uncertainty for states and countries that are clearly deviating from the power-law growth behavior like Switzerland, and New York is small, the uncertainty is extremely high for countries and states like Sweden and Pennsylvania that are still showing power law growth in dynamics. We return to this points below when we make forecasts and compare our results to those of the IHME model.

%One interesting observation is that the basic dynamics of a short period of exponential followed by power law growth seems to describe the infection dynamics of countries and states that have followed very different policies regarding social distancing. For example, countries like Sweden have followed a very limited social distancing policy whereas states like California have followed aggressive social distancing protocols. This suggests that the effects of social distancing, to the extent that they can be seen in the cumulative death and case data, are to change the maximum number of cases $N_{\max}$ and the duration of the infection (controlled by $\sigma$) rather than the overall shape of the curve.

\subsection{Data collapse indicate a universal dynamics for COVID-19 infections}

\begin{figure*}
	\centering
	\includegraphics[width=18cm]{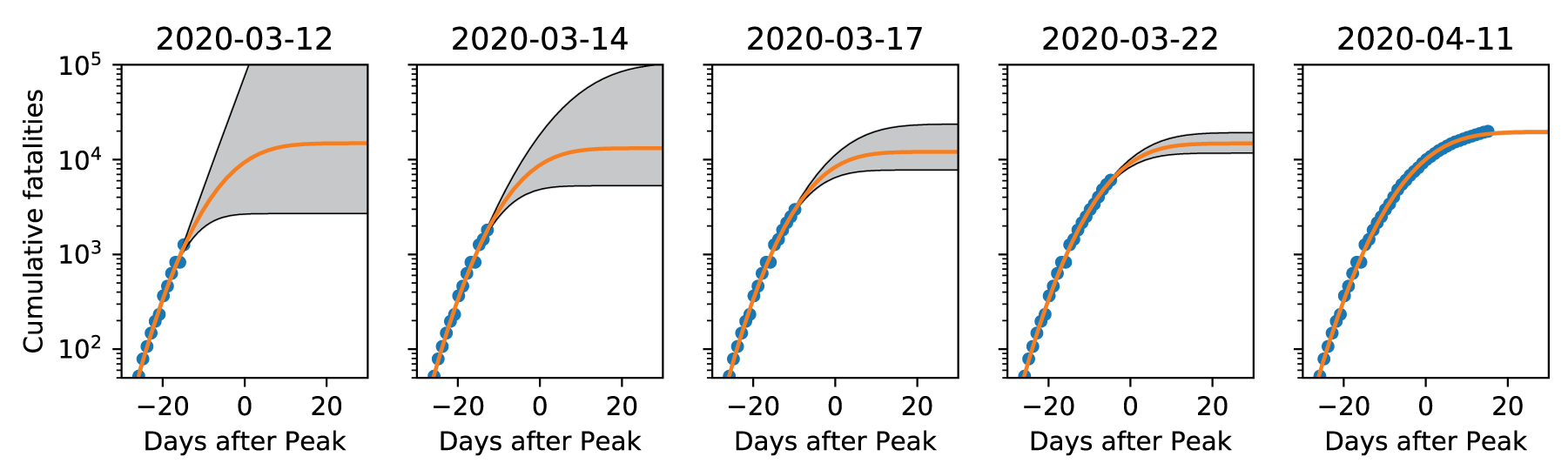}
	\caption{{\bf Prediction is difficult until a few days before peak}. Cumulative fatality data for Italy (blue points), along with best-fit to Eq. (\ref{eq:fittingfunction}) (orange lines) and confidence interval (gray region), using data collected before the date indicated at the top of each panel. Time axis is labeled as days before/after the estimated date of peak fatalities $t_h$. The first four columns show data collected 15, 13, 10 and 5 days before $t_h$, while the fifth is 15 days after $t_h$. Note that confidence intervals represent the \emph{minimal} level of uncertainty, under the assumption that the data is accurately described by Eq. (\ref{eq:fittingfunction}) plus random variations. Systematic deviations from this behavior caused by policy changes or failure of the fitting function at late times are not accounted for in this estimate.}
	\label{fig:forecast}
\end{figure*}

These observations prompted us to better understand the success of Eq. (\ref{eq:fittingfunction}) in fitting COVID-19 data across regions and countries. In particular, we wanted to ascertain if the reason that the erf function gives such good fits to curves across all regions and countries is because this is indicative of some underlying universal dynamics governing COVID-19 spreading. One powerful method for detecting universal dynamics that has its origin in the analysis of physical systems is the idea of a ``data collapse'' \cite{de1979scaling, newman1999monte}.  Namely, by rescaling the y-axis (cumulative number of deaths or confirmed cases $N$) and the x-axis (the time $t$ in days) it should be possible to make the curves for all regions and countries lie on each other, indicating the existence of a universal curve that describes infection dynamics.

The left hand panels of Figure \ref{fig:collapse}  show the cumulative deaths for countries and US states with at least 500 deaths. The right hand panels of the same figure show the same curves after we  have rescaled and shifted the time axis according to the function  $t \rightarrow (t-t_h)/\sigma$ and rescaled the y-axis according to $N \rightarrow N/N_{max}$. This scaling results in a striking collapse of the data with all the curves lying on top of each other. This is true for \emph{all regions} regardless of numerous details that differ between regions including population density, population size, social distancing prescriptions, the start time, duration, and severity of the COVID-19 outbreak -- though we emphasize that all countries shown in the top row had adopted some form of social distancing by the time of their fifth fatality. The last row of Figure 3 shows the same data collapse for all 49 countries and 30 states with at least 50 deaths on April 10th. Remarkably, the same universal curves seem to capture the disease dynamics for almost all regions, despite the numerous differences between countries and states in policy and composition.

The data collapse has numerous practical implications. First, it indicates that most of the complexities and place-specific contingencies of disease spread under most currently implemented forms of social distancing manifest themselves through just two parameters: a natural time scale $\sigma$ for the infection, and the total number of deaths/cumulative infections $N_{max}$.  Second, the collapse suggests that the primary effect of variations in policy and population characteristics is likely to modify these two parameters, especially $N_{max}$. Finally, it implies that if we can fit these two parameters directly we should be able to make reasonable forecasts for how many cases we expect and how long it will take the disease to spread assuming current policies (or something similar) continue. As shown in Fig. \ref{fig:sup3}, a similar data collapse is also observed if one uses the number of confirmed cases instead of the number of deaths.

\begin{figure*}[t]
	\centering
	\includegraphics[width=17cm]{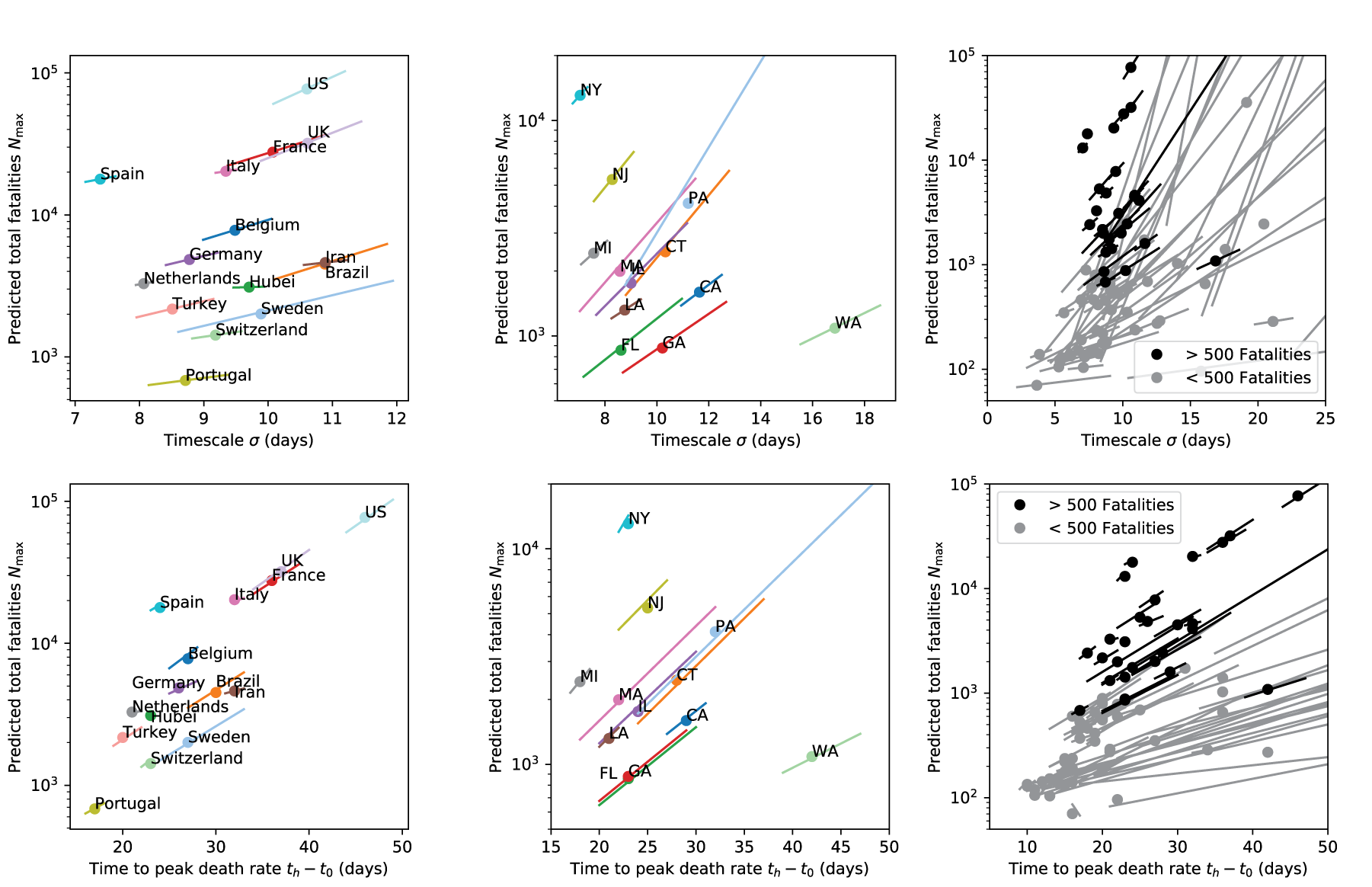}
	\caption{{\bf All regions show similar timescales}. Left: Predicted total number of fatalities $N_{\rm max}$ versus spreading timescale $\sigma$ and time to peak $t_h-t_0$ for countries with more than 500 fatalities as of April 15. Center: Same for all US states with more than 500 fatalities. Right: Same for all regions that qualified for fitting (including US states). The ends of the error bars in all panels lie at the joint upper and lower confidence bounds for the pair of parameters ($N_{\max}$ and $\sigma$ or $N_{\rm max}$ and $t_h-t_0$). Parameter combinations compatible with the data should fall close to this line, as explained in Fig. \ref{fig:errorbars} of the Methods. Note that most regions with fewer than 500 fatalities have extremely large error bars, in agreement with the discussion in Fig. \ref{fig:forecast} about the difficulty of inferring parameters until late in the epidemic progression. These large error bars do not always pass through the best-fit point since for these very poor fits the second dimension of the Fisher Information is no longer negligible.}
	\label{fig:params}
\end{figure*}

\subsection{Prediction is difficult until a few days before peak}

At first glance, the existence of universal dynamics would seem to allow for early and accurate forecasting under fixed policy conditions. But the confidence intervals for Pennsylvania and Massachusetts in Fig. \ref{fig:predall} indicate that a wide range of parameter values can provide excellent fits to the data even three weeks on from the beginning of the fatality curve. To gain a better understanding of when prediction becomes possible, we studied the model predictions for fatality dynamics in Italy, whose initial course of infection was nearly complete by April 15. We selected four time points before the time of peak fatalities and made predictions based on data collected before each time point, which were then compared with the actual outcome. Fig. \ref{fig:forecast} shows the results of this exercise, plotted on a log-linear scale.

We found that the behavior of the confidence interval can be categorized into three distinct regimes. In the initial exponential phase (first panel), the fit contains virtually no information about the time of the peak or the final number of fatalities. This can be understood by noting that in absence of any priors on parameters, unbounded exponential growth is compatible with our fitting function. In the subsequent power-law regime, the confidence interval becomes finite, but remains large (with a spread of greater than 15,000 deaths in the third panel). Finally, when the data departs from power law behavior, a few days before the time $t_h$ of peak deaths, the uncertainty shrinks to a small fraction of the predicted total number of fatalities. 

We conclude that even when social policies are fairly consistent, prediction based on historical data for a single region is possible only after the power law regime begins, and only becomes reliable a few days before the peak. 

\subsection{Forecasts from our universal dynamic model}

Our fitting procedure yields three parameters for each region/country: $N_{max}$, the predicted number of deaths, $\sigma$, which sets a natural time scale for the dynamics in days, and $t_{h}$ which indicates the expected time that new infections will peak. Since the errors in our fits are effectively one-dimensional (see Figure \ref{fig:errorbars} and Material Methods), we can also calculate an upper and lower bound on all three parameters, reflecting our uncertainty about predictions. Figure \ref{fig:params} shows a summary of our best fits to these three parameters with error bars. As can be seen, while $N_{\max}$ and $t_{h}$ are quite variable across regions and countries, likely reflecting different population characteristics and social policies, $\sigma$ is consistently around 8-11 days for almost regions and countries (the US state of Washington being a notable outlier).This may reflect the fact that the time scale $\sigma$ is a property of the virus itself rather than a property of the region in which the disease is spreading. This time scale is consistent with physiological data on recovery times after virus exposure \cite{bar2020sars, li2020substantial, he2020temporal}. Detailed predictions for all states and countries are provided in corresponding interactive notebook and CSV files on github.

\begin{figure*}[t]
	\centering
	\includegraphics[width=17cm]{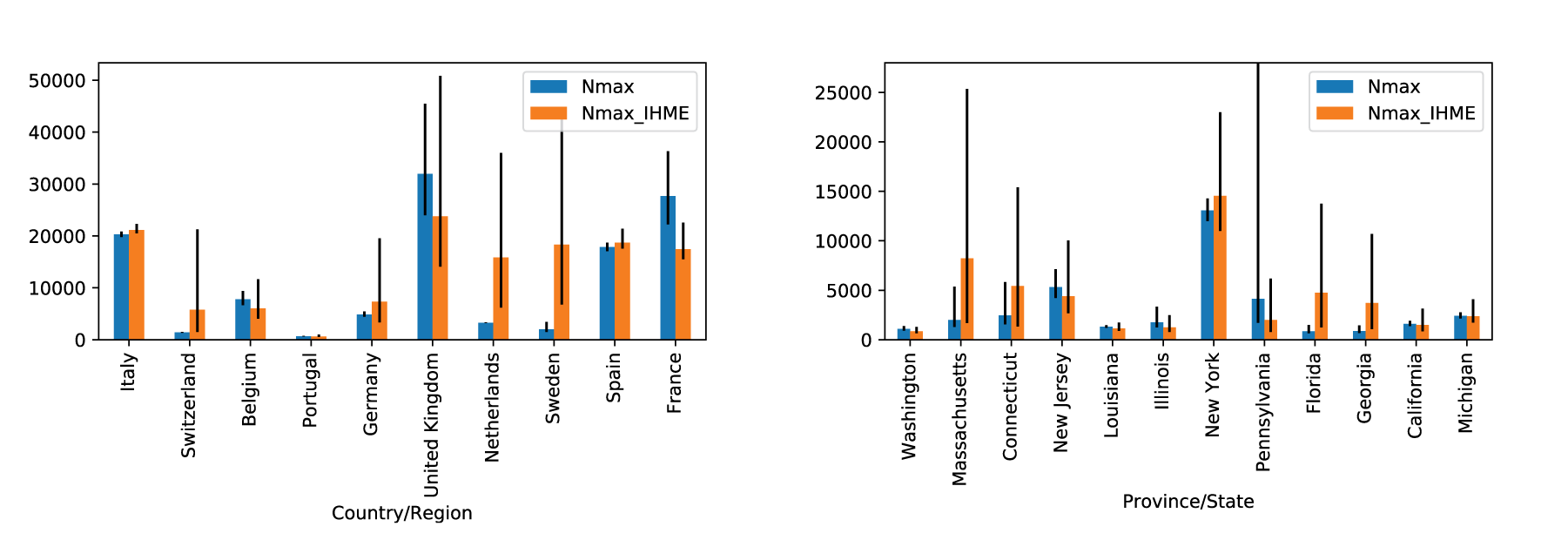}
	\caption{{\bf Comparing independent fits with globally linked IHME predictions}.  Our predictions (blue, Nmax) and IHME prediction (orange, Nmax\_IHME) for the final total number of fatalities in regions and states analyzed by IHME that had more than 500 fatalities as of April 15. Differences between the two predictions represent the effect of the linkage functions and prior expectations employed in the IHME forecasting pipeline.}
	\label{fig:IHME}
\end{figure*}

We also compared our forecasts to those from the IHME  \cite{covid2020forecasting}, as shown in Fig. \ref{fig:IHME}. Notice that for most states and countries the two models agree. However, for some countries (Sweden, Netherlands) and US states (Georgia and Massachusetts) there are significant discrepancies despite the fact that the two models use the same fitting function. This likely reflects the very different assumptions used in our fitting procedures. The IHME makes very specific assumptions about how social distancing affects the number of deaths, places strong global priors on parameters that are shared across regions and countries, and significantly constrains $\sigma$ using the Wuhan, China data. In contrast, our model is agnostic to any social distancing procedures or population characteristics and fits each region and country independently. For this reason, it reflects a more ``data-oriented'' approach that make no prior assumptions about COVID-19 disease dynamics, though it does assume that current social policies will continue until the fitting function saturates. It will be interesting to see which of these fitting procedures yields better long term predictions where there is disagreement between the two forecasts.

\subsection{Assessing the effects of social distancing}

One of the most surprising aspects of our analysis is that regions and countries with very different social distancing policies all fall on the same universal curve after rescaling. For this reason, we sought to better understand the effect of social distancing on COVID-19 dynamics. To do so, we made use of a country-specific  social stringency index compiled by Oxford COVID-19 Government Response Tracker (OxCGRT) \cite{OxfordCOVID19}. The OxCGRT was created to provide ``a systematic way to track the stringency of government responses to COVID-19 across countries and time.'' Starting on Jan 1, 2020, on each day the OxCGRT assigned a stringency score to each country between 0 and 100 that seeks to summarize policies such as school closures, travel bans, etc. with 0 being least stringent and 100 corresponding to maximum stringency.

\begin{figure*}[t]
	\centering
	\includegraphics[width=17cm]{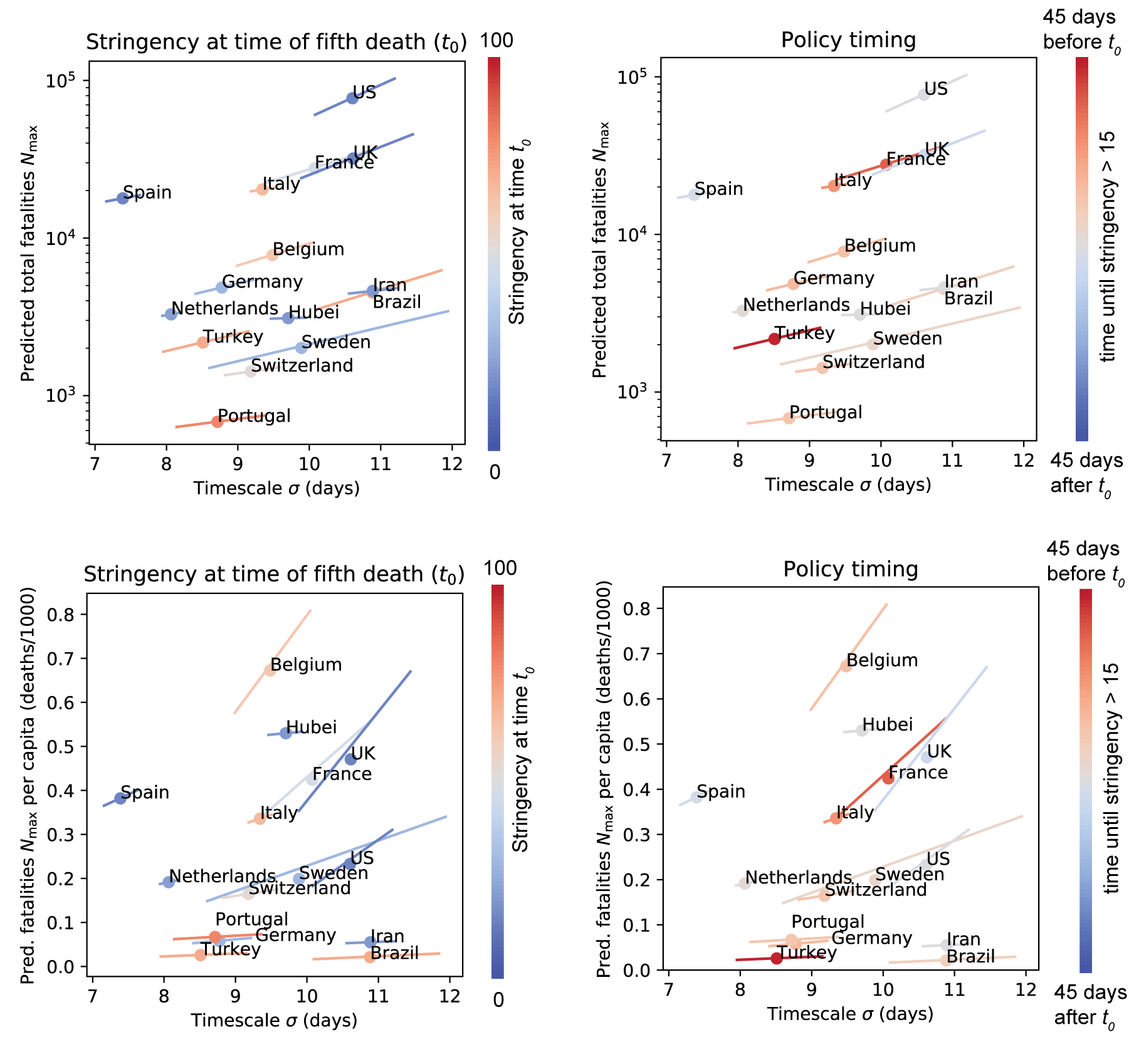}
	\caption{{\bf Effects of social distancing and travel restrictions on parameter values}.  Predicted total number of fatalities $N_{\rm max}$ (absolute and per-capita) versus timescale $\sigma$ for countries with more than 500 fatalities, with data points colored by stringency of social distancing plus travel restrictions as measured by OxCGRT \cite{OxfordCOVID19}. Left: Stringency at the time $t_0$ of the fifth fatality. Right: Number of days from $t_0$ to the time when the stringency score reached 15.  Note that all countries shown had implemented at least this level of restriction (which requires at least two of the seven policies included in the score) within a few days of the fifth death.  }
	\label{fig:sd}
\end{figure*}

As discussed above,  the existence of universal dynamics means that each country can be characterized by two parameters: $N_{max}$, the predicted number of deaths, and $\sigma$, a time scale governing the dynamics. The top row of Figure \ref{fig:sd} shows scatter plots of these two quantities, colored by stringency at the time of the fifth death denoted by $t_0$  (left panel) and the time, measured from $t_0$, it took a country to implement at least two social distancing measures corresponding to an  OxCGRT index of at least 15 (right panel). Note that red corresponds to high stringency and blue to low stringency. As is clear, all countries have implemented social distancing measures, but there is no obvious correlation between the exact level of social distancing and $N_{\rm max}$.  The bottom row shows identical plots except the y-axis now corresponds to the predicted per-capita death rates (predicted deaths per 1000 people) instead of the total number of deaths. Once again, there is no obvious relationship between the OxCGRT index and predicted per capita death rates.

This suggests that discerning the effects of social policy on COVID-19 dynamics may be quite subtle, and great care should be taken when making causal arguments relating policy choices to outcomes. Alternatively, it may reflect the fact that different regions and countries have very different reporting standards, making it difficult to compare variations in social policy and deaths across regions and countries.

\subsection{Origin of universal dynamics for disease spreading}

 \begin{figure*}[t]
\centering
\includegraphics[width=17cm]{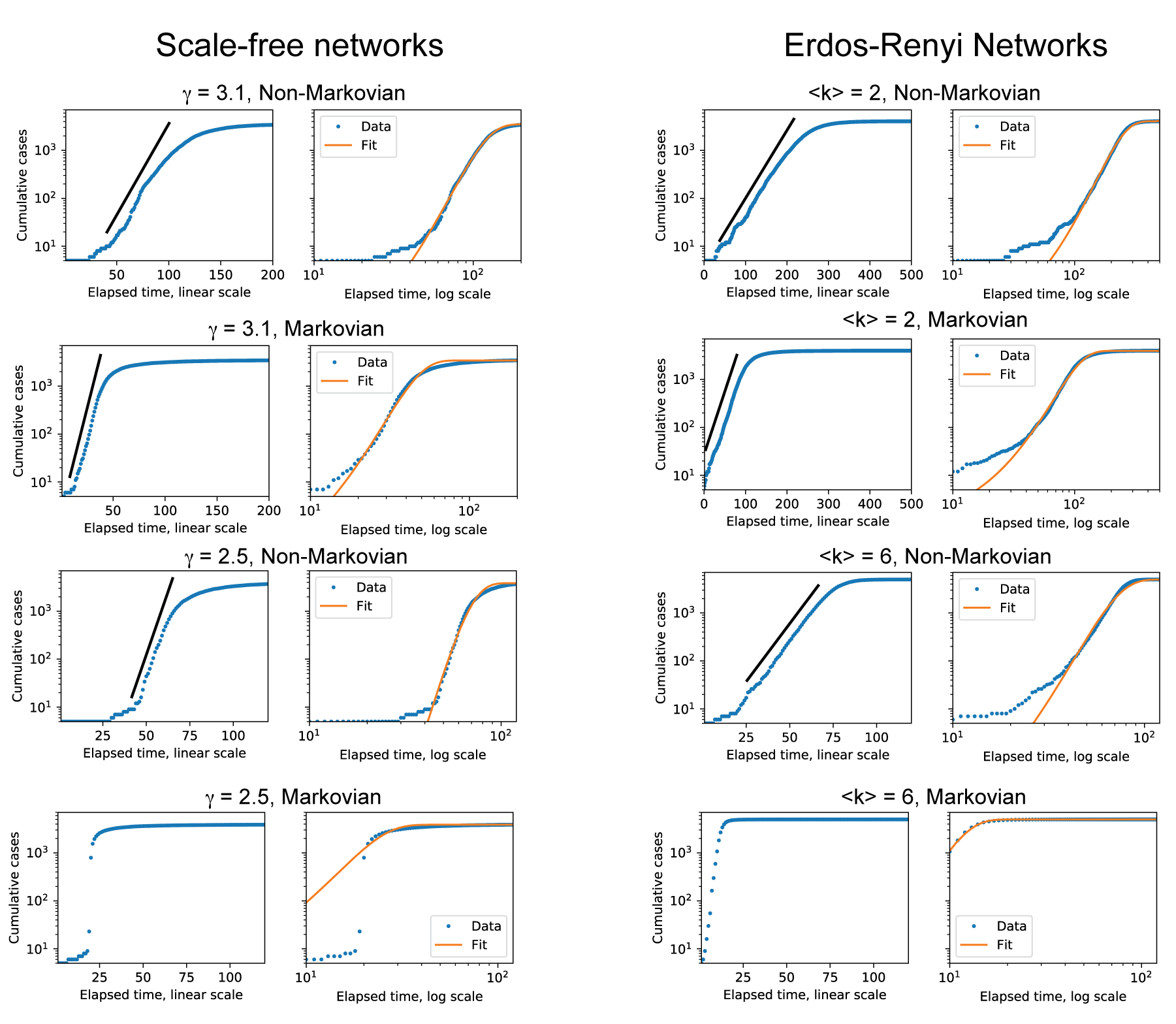}
\caption{{\bf Scale-free and Erd\"{o}s-R\'{e}nyi random networks can reproduce observed dynamics.} Simulations of virus spreading on scale-free and Erd\"{o}s-R\'{e}nyi random networks with 5,000 nodes were performed as described in the main text and methods. Scale-free networks were sampled with degree distribution $p(k) \propto 1/k^\gamma$, while Erd\"{o}s-R\'{e}nyi networks were sampled with mean degree $\langle k\rangle$, shown in each panel title. The time required for a node to infect its neighbor is sampled from a Gamma distribution with mean $\mu_G = 50$. Non-Markovian dynamics were generated by choosing standard deviation $\sigma_G = \mu_G/2 = 25$, while Markovian dynamics were generated by choosing $\sigma_G = \mu_G = 50$ (because the Gamma distribution reduces to the exponential distribution when $\mu_G = \sigma_G$). Left column is log-linear plot while right is log-log. Black line is a straight line to guide the eye. Orange line is fit to the cumulative normal given in Eq. (\ref{eq:fittingfunction}). The non-Markovian dynamics produce excellent fits (after the initial stochastic regime) for $\gamma > 3$ and for small $\langle k\rangle$, while the fit quality degrades for Markovian dynamics or for smaller $\gamma$ or large $\langle k\rangle$.}
\label{fig:scalefree}
\end{figure*}

As noted above, power-law spreading at early times has been observed in several large-scale epidemics, including HIV/AIDS and the 2014 Ebola outbreak \cite{viboud2016generalized}. This initial sub-exponential dynamics implies that the average number of new infections generated by each infectious individual begins decreasing even before a significant percentage of the population has been exposed to the disease. This surprising behavior is incompatible with many  epidemiological models based on differential equations (e.g.  SIR type models with or without population structure, compartments, etc.), and also with a variety of individual-based network models in both homogeneous and inhomogeneous networks \cite{newman2006structure, barthelemy2005dynamical, barrat2008dynamical,pastor2015epidemic,pastor2015epidemic} (see \cite{vazquez2006polynomial} for notable exception). A number of mechanisms have been proposed to account for this phenomenon, including time-varying parameters due to behavioral changes, and the constraints on network topology imposed by two-dimensional spatial structure \cite{chowell2016mathematical}. 

%Disease spreading on social networks is an extremely well studied problem \cite{newman2006structure, barrat2008dynamical, pastor2015epidemic,  barabasi2016network}, and a complete review of this extensive and interesting literature is beyond the scope of this work and our abilities. 

Mechanisms involving behavioral change are particularly plausible for the data shown here, where all regions analyzed have implemented some form of social distancing. Here, we tested an alternative hypothesis: the universal curves may be due to non-Markovian dynamics and significant  individual-level heterogeneity in the time it takes an individual to transmit the virus to their neighbors on a social interaction network.  Consistent with this idea, recent evidence suggests that SARS2-COVID19 may exhibit considerable inter-individual heterogeneity \cite{bar2020sars, li2020substantial, he2020temporal}. To test this hypothesis, we performed simulations of disease spreading on social networks on scale free networks where the degree distribution $p(k)$ take the form $p(k) = {1 \over k^\gamma}$. Such power-law distributed graphs have nodes with large degree distributions and have been argued to be more realistic approximation to real social interactions  \cite{newman2006structure, pastor2015epidemic,  barabasi2016network}. We also performed simulations on random  Erd\"{o}s-R\'{e}nyi graphs with different mean degree.

 In each network, $N_0$  nodes (usually $5$ in our simulation) were initially designated infected. An infected node infects a susceptible neighbor node at a time $T_g$ after being infected, where $T_g$ is edge-specific random variable drawn from a Gamma distribution with mean $\mu_G$ and standard deviation $\sigma_G$. This procedure is iterated  until no new nodes can be infected (see Materials and Methods and corresponding Jupyter notebooks on our Github repository for details).  Note that when $\mu_G =\sigma_G$ the Gamma distribution reduces to the exponential distribution and the dynamics is Markovian (i.e. memoryless). However, if $\mu_G \neq \sigma_G$ then the dynamics is Non-Markovian. Non-Markovian dynamics introduces additional time-scales into the problem and hence is much less studied than its Markovian counterpart \cite{pastor2015epidemic}.
 
 Figure \ref{fig:scalefree} shows the cumulative number of infected nodes as a function of time for simulated graphs with $N=5,000$ nodes. Simulations were performed for scale-free networks with $\gamma=2.5,3.1$ and  Erd\"{o}s-R\'{e}nyi random networks with mean degree $\langle k\rangle =2,6$, with both Markovian ($\mu_G =\sigma_G=50$) and non-Markovian ($\mu_G=50, \sigma_G=25$) dynamics. The non-Markovian dynamics produce excellent fits (after the initial stochastic regime) for $\gamma > 3$ and for small $\langle k\rangle$, while the fit quality degrades for Markovian dynamics or for small $\gamma$ or large $\langle k\rangle$. This suggests that non-Markovian dynamics and variability among individuals in the time they take to infect neighbors maybe an important feature controlling disease spreading in these networks.  
%%%%%%%%%%%%%%%%%%%%%
 
\section{Discussion}

In this paper, we have used timeseries data from the Johns Hopkins University Coronavirus Resource Center on the cumulative number of confirmed cases and deaths due to COVID-19 to build a data-driven model of infection dynamics under social distancing.  We find that the COVID-19 pandemic  in all regions with sufficient data can be described by a universal curve where the cumulative numbers of both infections and deaths quickly cross over from exponential growth at early times  to a longer period of power law growth, before eventually slowing. In agreement with a recent statistical forecasting model by the IHME, we show that this dynamics is well described by the cumulative distribution function of the normal distribution. Surprisingly, we find that the despite the enormous variation in region/state/country characteristics and social policy, the infection dynamics across all regions/countries analyzed can be characterized by two quantities:  the predicted number of deaths/confirmed cases in a region $N_{max}$ and a scale $\sigma$ that sets the natural timescale of infection in a region. Rescaling time and number of deaths/confirmed cases by these two quantities resulted in a remarkable data collapse of infection dynamics across regions, providing strong evidence for our claim of universal dynamics under current social distancing policies.

Our analysis also suggests that, for countries and regions that have crossed the 50 death threshold, the predictive power of any statistical model based on tracking deaths or cases is likely to be extremely limited until a few days before the infection peaks. This suggests that it is very difficult to make predictions before curves start flattening. However, after the infection peaks, the uncertainty drops dramatically. This suggests that while it is very hard to make predictions early on, at later times it is possible to predict both the severity and duration of the infections with much more confidence. This uncertainty is present even though social policies may not be changing much. 

We have used our statistical model to make forecasts for regions, states, and countries where we have enough data. Importantly, our forecasts assume that there will be no dramatic shifts in social distancing policies. While we use the same fitting function as the IHME, our fitting procedures have crucial differences in the assumptions they make.  Crucially, the IHME fitting procedure makes specific assumptions about the effects of social distancing  and places strong priors on fitting parameters. In contrast, our model is agnostic to any social distancing procedures or population characteristics and fits each region and country independently. Nonetheless, we are able to obtain very good fits to the data.

While many of the predictions between the two models show good agreement, there are some prominent disparities. For example using data up to April 15th for the US state of Massachussets, the IHME model predicts that there will be more than 8,200 deaths (with confidence intervals between 1,680 and 25,347) and the infection will peak on May 5th. In contrast, our model predicts 1,993 deaths (with confidence intervals between 1,302 and 5,366), with a peak time of April 14th. This likely reflects the fact that the IHME model imposes a large penalty on Massachusetts for its somewhat less stringent social distancing policies. A similar explanation likely holds for the case of Florida. It will be interesting to compare and contrast these two distinct approaches to fitting to better understand how to best forecast COVID-19 disease dynamics assuming that social policies continue as is.

We also provided a preliminary exploration of how social distancing policies affect the parameters of the universal dynamics. To do so, we made use of the country-specific  social stringency index compiled by Oxford COVID-19 Government Response Tracker \cite{OxfordCOVID19}. While there were some anecdotal features in the per-capita fatalities of the most lax and most stringent countries, much more work needs to be done to adequately understand how different stringencies of social distancing affect disease spreading dynamics. One important caveat of our analysis of the effects of social distancing is that our statistical modeling approach is geared towards countries with relatively large death totals. Our model has little to say about countries like South Korea (about 230 deaths on April 15th) that have used extensive testing and social distancing to successfully contain the COVID-19 pandemic. Much more work will have to be done to understand this in greater detail.

We emphasize that our statistical model is neither capable of, nor designed to, understand what will happen if policies change substantially. All of the countries we have analyzed have adopted significant social distancing protocols. Since we do not explicitly incorporate the effects of policy in our fits, we have no way of asking about what will happen if these policies change significantly. Nonetheless, there seem to be some general lessons to be learned. First, despite all the variation across regions and countries, cases and deaths seem to quickly cross over from exponential to power law growth. Similar behavior was observed in other epidemics including HIV/AIDS and the 2014 Ebola outbreak \cite{viboud2016generalized,chowell2016mathematical}. This suggests that it is useful to plot all data in both log-linear and log-log scales. Second, it cautions against making extrapolations far into the future based on exponential growth, since power law growth seems to be quite consistent and generic.

Finally, to better understand the origin of these dynamics, we performed simulations of disease spreading on social interaction networks. We focused on dynamics where the time it takes an individual to infect their neighbors follows a non-Markovian waiting time distribution. Furthermore, we assumed that individuals could have significant heterogeneity in their infection times. With these assumptions, we were able to reproduce the universal dynamics observed in real COVID-19 data using our simulations. While much more work needs to be done to understand the origin of this universal dynamics, these simulations suggest that the combination of non-Markovian dynamics  and significant heterogeneity among individual transmission times may be crucial factors governing disease spreading dynamics. %These observations are consistent with recent studies suggest that there is significant inter-individual variability in COVID-19 infections  \cite{bar2020sars, li2020substantial, he2020temporal}.

\subsection{Practical implications and lessons for policy}
We conclude by discussing some practical implications and lessons from our work. 

\begin{itemize}
\item  {\bf Slowing down in the rate of new cases/deaths in the early weeks may not indicate progress}. We found that are after a short period of exponential growth, the number of confirmed cases and deaths quickly transitioned to power law growth (about 10 days after the fifth confirmed death). We observed this basic phenomena in regions with extremely different population characteristics and social policies. Since power law growth is much slower than exponential growth, this serves as a strong caution against viewing a slowing down in the increase of deaths/confirmed cases in the first few weeks as progress in slowing down COVID-19 spread or  indicating the success of any particular policy measures or actions.  This is  consistent with observations in several other large-scale epidemics, including HIV/AIDS and the 2014 Ebola outbreak \cite{viboud2016generalized}.

\item {\bf Plot deaths/confirmed cases on log-log plot rather than log-linear plots.} Since infection dynamics quickly transition over to power law-like dynamics after two weeks, to assess whether the number of cases/deaths is flattening it is better to plot data on a log-log scale (e.g. log deaths vs log time). In such plots, the number of deaths/confirmed cases will fall on a straight line whose slope indicates the power-law exponent. 

\item{\bf Flattening should be assessed by deviation from straight lines on log-log plots.} The end of the power-law growth regime and a slow-down in new cases is indicated by deviations from a straight line on log-log plots. This is likely to be a much better indication of when infections have peaked than other heuristics based on comparing to exponential growth (at least if social policies are fairly consistent).

\item{\bf Accurately predicting the expected number of deaths or infection duration is very difficult until a few days before the peak. } In contrast, after infections peak, the uncertainty drops quickly allowing for much better estimates of infection duration and predicted number of deaths/confirmed cases. Both these facts should be considered when using forecasts to implement social policies such as the end of social distancing.

\item{\bf Standard epidemiological models based on measuring $R_0$ may be missing crucial aspects of disease spreading}. Classical disease models based on differential equations often characterize disease dynamics using a single parameter $R_0$, the average number of new infections from an infected individual. However, the dynamics produced by such models seems to differ significantly from the universal COVID-19 dynamics that we observe.

\item{\bf Projections about future outbreaks based on exponential growth may be inaccurate.} One consistent trend observed in our data is that the exponential growth phase quickly gives rise to power law growth. This seems to be true even for countries like Sweden whose social distancing efforts, though put into place early in the epidemic progression, were relatively mild (canceling public events and carrying out a public information campaign). If this trend holds true during future outbreaks after social distancing policies are weakened (e.g. relaxed to those of Sweden),  then extrapolations based on measuring $R_0$ and assuming exponential growth are likely to severely overestimate the number of deaths/infections. 
\end{itemize}

%%%%%%%%%%%%%%%%

\section{Materials and Methods}
\subsection{Data acquisition and preprocessing}

We obtained global and US timeseries of fatalities and confirmed cases from the Johns Hopkins CSSE COVID-19 Github repository, \url{https://github.com/CSSEGISandData/COVID-19}. For each region (country/province or US state), we estimated the initial times $t_0$ of the fatality and confirmed cases curves to be the first day the cumulative number (deaths or cases, respectively), exceeds 5. The time $t_0$ is only used for constructing the time axis of the log-log plots and for estimating the time interval $t_h-t_0$ in Figure \ref{fig:params}.

For fitting, prediction and uncertainty estimation, we discarded fatality data for times before 50 cumulative fatalities were reached, and confirmed cases data for times before 500 cumulative cases. This allowed us to focus on the universal behavior at later times, without the influence of the stochastic effects that dominate the very early phase of the infection. We obtained fits and confidence intervals for regions with at least six data points remaining after this truncation.

IHME predictions were downloaded from \url{http://www.healthdata.org/covid/data-downloads} on April 16, 2020. Data from Oxford COVID-19 Government Response Tracker index was downloaded from
\url{https://www.bsg.ox.ac.uk/research/research-projects/coronavirus-government-response-tracker} \cite{OxfordCOVID19}. 

\subsection{Model fitting and uncertainty estimation}
As stated in the main text, we took the predicted cumulative number $\hat{N}(t)$ of fatalities (and cases for the supplementary figures) to follow the functional form
\begin{align}
\hat{N}(t) = N_{\rm max}\left[\frac{1}{2}+{\rm erf}\left(\frac{t-t_h}{\sqrt{2}\sigma}   \right) \right] = N_{\rm max}\Phi\left(\frac{t-t_h}{\sigma}\right)
\label{eq:hatN}
\end{align}
where 
\begin{align}
\mathrm{erf}(t) &={1 \over \sqrt{\pi}} \int_0^t e^{-x^2}\,dx\\
\end{align}
is the Gaussian error function and
\begin{align}
\Phi(t) &= {1 \over \sqrt{2\pi}} \int_{-\infty}^t e^{-x^2/2}\,dx
\end{align}
is the cumulative normal distribution. This model has three fit parameters: the half-max time $t_h$, the width $\sigma$, and the final cumulative total $N_{\rm max}$. We performed separate fits for each geographical region in the JH database, allowing all three parameters to vary independently from region to region. 

\subsubsection{Error model and maximum likelihood estimation}
To obtain a maximum-likelihood estimate for these three parameters and corresponding confidence intervals, we used a simple multiplicative noise model
\begin{align}
N(t) = \hat{N}(t)e^{\xi(t)}
\end{align}
where $\xi(t)$ is Gaussian white noise with standard deviation $\sigma_\xi$. The probability of observing a set of cumulative counts $N_i = N(t_i)$ at times $t_i$ ($i = 1,2\dots M$), given a parameter set $\theta = (t_h,N_{\rm max},\sigma)$ under this model is
\begin{align}
p(N_i|\theta) \propto {\rm exp}\left(-\frac{1}{2\sigma_\xi^2}\sum_{i=1}^M [\log N_i - \log \hat{N}_i(\theta)]^2\right) \label{eq:N}
\end{align}
where $\hat{N}_i(\theta)$ is the prediction at time $t_i$ using the parameter set $\theta$. Now by Bayes' rule, the probability that $\theta$ is the true set of parameters, given the observed data, is
\begin{align}
p(\theta|N_i) \propto p(N_i|\theta) p(\theta)
\end{align}
where $p(\theta)$ represents our prior expectations for the parameter values. In this analysis, we used a uniform prior, so $p(\theta)$ is a constant. The ``best-fit'' parameter set $\hat{\theta}$ is defined as the one that maximizes the posterior probability $p(\theta|N_i)$ for the given data $N_i$. For uniform priors, this is the same as Maximum Likelihood.

We calculated $\hat{\theta}$ by minimizing the cost function
\begin{align}
C(\theta) = \frac{1}{2} \sum_{i=1}^M [\log N_i - \log \hat{N}_i(\theta)]^2
\end{align}
over $\theta = (t_h,\log N_{\rm max},\sigma)$ using the Nelder-Mead method as implemented in the Python package SciPy \cite{scipy}, where $\hat{N}_i = \hat{N}(t_i)$ is given by equation (\ref{eq:hatN}) above. 

\begin{figure}
	\includegraphics[width=8cm]{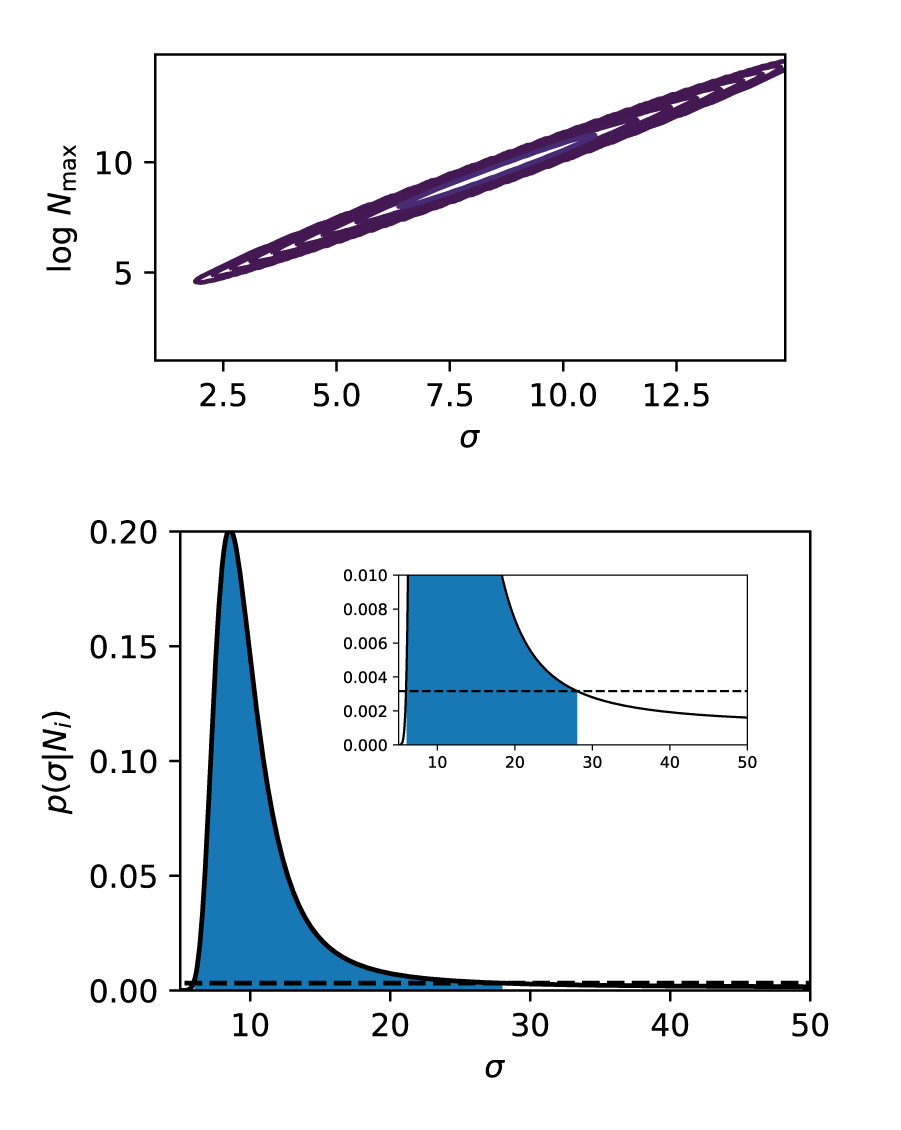}
	\caption{{\bf Uncertainty estimation}. Left: Contours of the Gaussian approximation to $p(\theta|N_i)$, after marginalizing over $t_h$. The values of $\sigma$ and $N_{\rm max}$ are highly correlated, such that the fit is effectively one-dimensional. Contour lines are inverse powers of 10 ($10^{-1},10^{-2},\dots,10^{-9}$) times the height of the peak $p(\hat{\theta}|N_i)$. Right: Probability $p(\sigma|N_i)$ obtained by sweeping $\sigma$ and obtaining the best-fit values of $t_h$ and $N_{\rm max}$ at each value of $\sigma$. Shaded area contains 95\% of the probability, and is bounded by the upper and lower confidence bounds on $\sigma$. Both panels are generated using Italy fatality data, including only data up to 15 days before estimated peak (i.e, up to March 12, first column in Fig. \ref{fig:forecast}), so that uncertainty is still large. }
	\label{fig:errorbars}
\end{figure}

\subsubsection{Initial conditions for optimization}
Due to the highly nonlinear nature of this fit, it important to supply the optimizer with good initial conditions. We obtain an initial estimate for $t_h$ and $\sigma$ using the mathematical relationship between the mean and variance of a truncated normal distribution with the full distribution. 

First, we compute the daily number of new fatalities/cases $n(t) = dN/dt$. If the cumulative number $N$ is a cumulative normal, the distribution $n(t)$ of death/infection times is normal. If the epidemic has not yet concluded, however, the distribution is truncated at the latest observation time $t_f$. One can show that the mean $\langle t\rangle$ and variance ${\rm var}(t)$ of the death/infection times before $t_f$ are related to the mean $t_h$ and variance $\sigma^2$ of the full distribution by \cite{barr1999mean}:
\begin{align}
\langle t \rangle &= -\sigma c\left( -\frac{t_f-t_h}{\sigma}\right)e^{-\frac{(t_f-t_h)^2}{2\sigma^2}} + t_h\\
{\rm var}(t) &= \sigma^2 c\left( -\frac{t_f-t_h}{\sigma}\right)\bigg\{\sqrt{\frac{\pi}{2}}\bigg[1\nonumber\\
&+\frac{t_f-t_h}{|t_f-t_h|} C_3\left(\left(\frac{t_f-t_h}{\sigma}\right)^2\right)\bigg] \nonumber\\
&- c\left( -\frac{t_f-t_h}{\sigma}\right)e^{-\frac{(t_f-t_h)^2}{2\sigma^2}}\bigg\}
\end{align}
where
\begin{align}
c(t) &= \frac{1}{\sqrt{2\pi}[1-\Phi(t)]}\\
C_3(t) &= \int_{t^2}^\infty \frac{u^{1/2}e^{-u/2}}{2^{3/2} \Gamma(3/2)}du
\end{align}
and $\Phi(t)$ is the cumulative distribution of the unit normal. 

For each region, we obtained initial estimates of $t_h$ and $\sigma$ by solving these equations, using the current sample mean $\langle t\rangle$ and variance ${\rm var}(t)$ of the fatality/infection times. We then estimated $N_{\rm max}$ as $N_{\rm max} = N(t_f)/\Phi((t_f-t_h)/\sigma)$.

These estimates were used as initial conditions for our non-linear solver.

\subsubsection{Estimation of confidence bounds}
To obtain confidence bounds, we estimated $\sigma_\xi$ using the fact that the average cost $\langle C(\theta) \rangle = \frac{M}{2 }\sigma_\xi^2$ under the noise model given in Eq. (\ref{eq:N}). We thus approximated 
\begin{align}
\sigma_\xi^2 \approx \frac{2}{M}C(\hat{\theta}).
\end{align}
A Taylor expansion of $p(\theta|N_i)$ about $\hat{\theta}$ yields a Gaussian distribution 
\begin{align}
p(\theta|N_i) &\propto \exp\left( -\frac{C(\theta)}{\sigma_\xi^2}\right) \\
&\approx \exp\bigg[-\frac{1}{\sigma_\xi^2}\bigg(C(\hat{\theta}) \nonumber\\
&+ \frac{1}{2}\sum_{i=1}^3\sum_{j=1}^3 \frac{\partial^2 C}{\partial \theta_i\partial \theta_j} (\theta_i-\hat{\theta}_i)(\theta_j-\hat{\theta}_j)\bigg)\bigg],
\end{align}
with covariance matrix equal to the matrix inverse of the Hessian $\frac{\partial^2 C}{\partial \theta_i\partial \theta_j}$. The spectrum of this covariance matrix exhibits a ``sloppy mode'' for most of the regions analyzed when uncertainties are large (in the exponential or power law regime), with one eigenvector at least an order of magnitude larger than the other two. This means that the model uncertainty is effectively one-dimensional, and we can estimate the uncertainties by varying just one parameter, while fixing the other two by minimizing the cost $C$ at each value of the varied parameter. We chose to vary $\sigma$, and thus obtained a univariate probability distribution 
\begin{align}
p(\sigma|N_i) \propto e^{-C(\sigma)/\sigma_\xi^2}
\end{align}
where $C(\sigma)$ is the minimal value of $C$ at the given value of $\sigma$. By evaluating this function over a large range of $\sigma$ values, we normalize the distribution and compute the value $C_{95}$ for which 95 percent of the probability has $C\leq C_{95}$. The upper and lower confidence bounds for $\sigma$ are the boundaries of the region satisfying $C(\sigma)\leq C_{95}$, and the other two parameters are found by minimizing $C(\theta)$ at each fixed value of $\sigma$, as described above.

\subsection{Simulations of disease spreading on social interaction with heterogeneous waiting times}

We briefly summarize our simulation procedure. We simulated a generalized SIR model on a social interaction network $G$ with both Markovian and non-Markovian waiting times for infection.  Initially, $N_0$  nodes (usually $5$ in our simulation) are designated infected. An infected node infects a susceptible neighbor node at a time $T_g$ after being infected, where $T_g$ is an edge-specific random variable chosen from the distribution governing generation times for infections.

This procedure is iterated over multiple generations until no new nodes can be infected (i.e. all the nodes in the connected components of infected nodes are also infected). In our simulation, once a node is infected it cannot be reinfected either because the node joins the recovered population and is no longer susceptible or because the node dies. Our simulation makes no distinction between these possibilities and simply counts cumulative number of infected nodes as a function of time. We choose $T_g$ to be a Gamma distributed random variable  with mean $\mu_G$ and standard deviation $\sigma_G$. This assures that $T_g>0$. This differs from most simulations where generation times are assumed to be memoryless and follow an exponential distribution. Memoryless dynamics can also be implemented in this framework by choosing $\sigma_G = \mu_G$, since this parameter choice makes the Gamma distribution identical to the exponential distribution.

To make the plots in the text, we generated single instances of social interaction graphs $G$ with $N$ nodes using the NetworkX python package \cite{hagberg2008exploring}. Typically, in our simulations we chose $N=5,000$. We then ran the infection simulations as described above. We fit the cumulative number of infected nodes as a function of time using the same procedure as was used for the data (described above).

 In this paper, we focused primarily on scale-free and  Erd\"{o}s-R\'{e}nyi random networks generated using the NetworkX package. These networks are each characterized by a single parameter: the exponent $\gamma$ and the mean degree $\langle k\rangle$, respectively. 
 %scale free networks where the degree distribution $p(k)$ take the form of a power law:
%\be
%p(k) = {1 \over k^\gamma}.
%\ee
%Such power-law distributed graphs have nodes with large degree distributions and have been argued to be more realistic approximation to real social interactions  \cite{newman2006structure, pastor2015epidemic,  barabasi2016network}. We  also performed additional simulations on  Erdos-Reyni graphs which were also generated using  the NetworkX package.  

All code is available on our Github project page \url{https://github.com/Emergent-Behaviors-in-Biology/covid19}.

\section{Acknowledgments}
We would like to thank Nicole Aschoff, Arup Chakraborty, and Alex Gourevitch for useful discussions/comments  on the manuscript.
P.M. would like to thank the members of his BU Biophysics class for the discussions, and questions that motivated this work. 
We would also like to thank  Marin Bukov, Wenping Cui, Charles Fisher, Josh Goldford, Arvind Murugan, Dries Sels and Rob Phillips  for useful critical comments on the manuscript.
The work was supported by NIH NIGMS grant 1R35GM119461,  Simons Investigator in the Mathematical Modeling of Living Systems (MMLS) to PM.

\bibliography{refs_corona.bib}

\onecolumngrid

\newpage

\appendix

\begin{figure*}
    \centering
    \includegraphics[width=15cm]{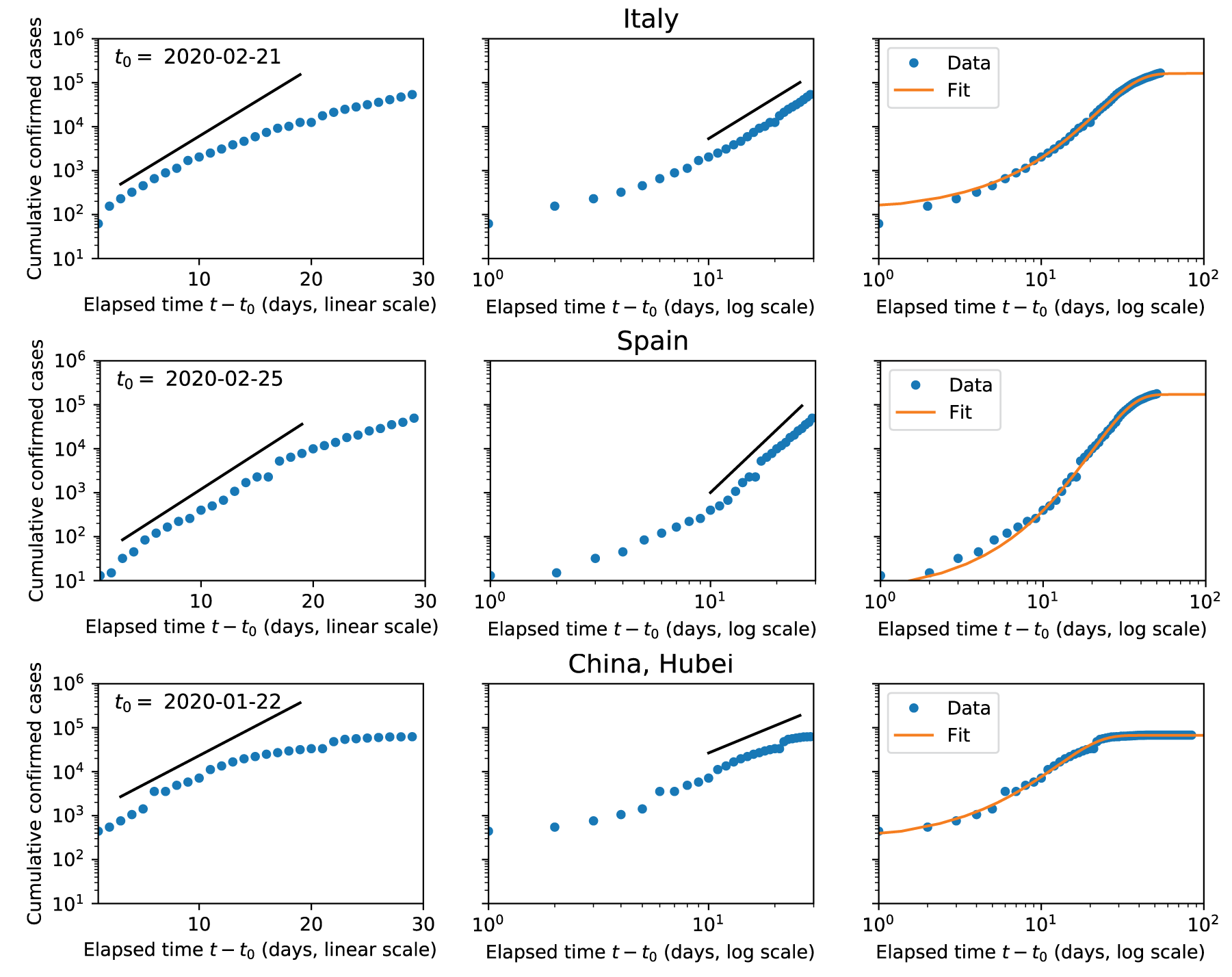}
    \caption{As Fig. \ref{fig:CIS}, but with cumulative case data instead of fatality data.}
    \label{fig:sup1}
\end{figure*}

\begin{figure*}
    \centering
    \includegraphics[width=15cm]{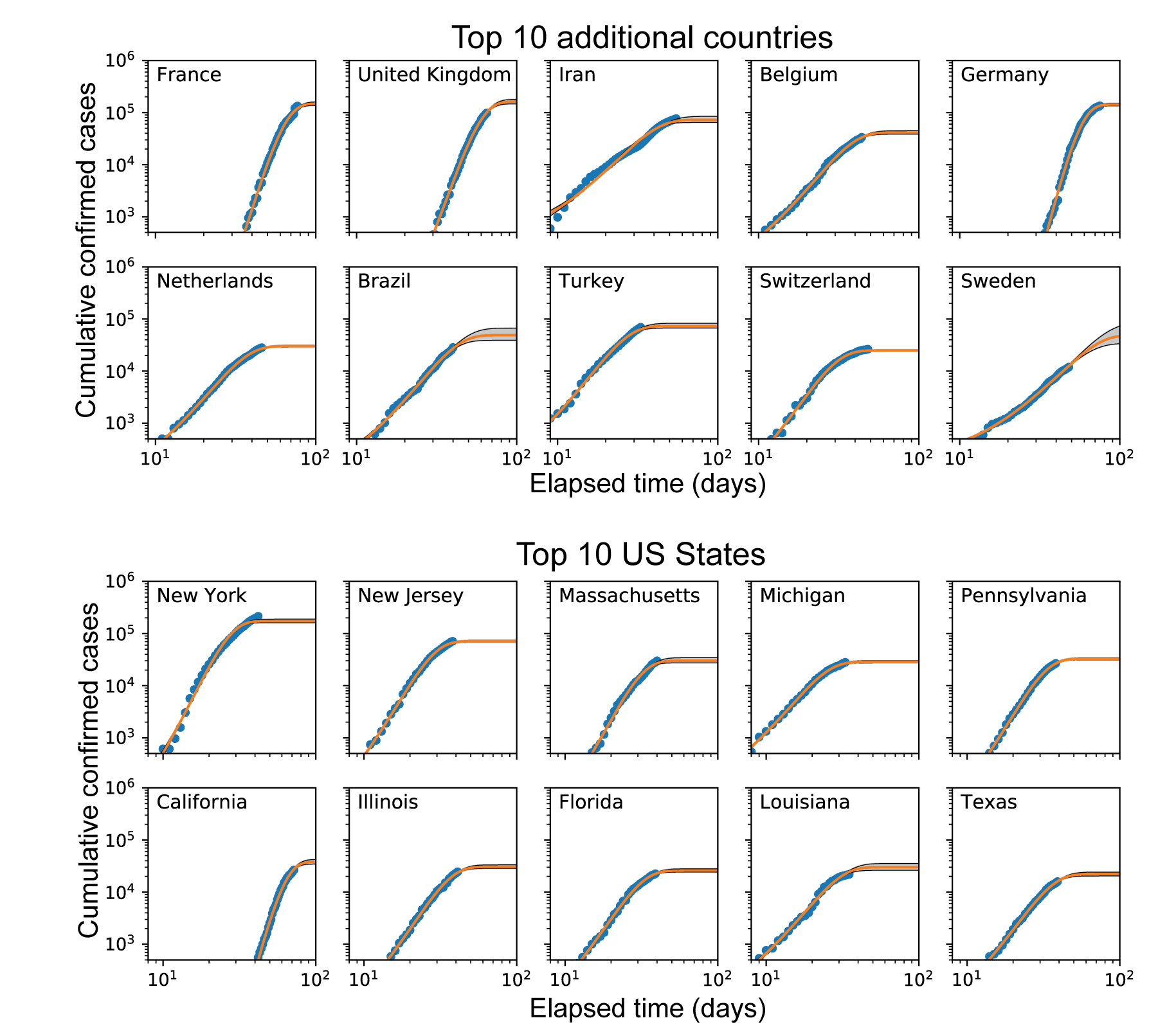}
    \caption{As Fig. \ref{fig:predall}, but with cumulative case data instead of fatality data.}
    \label{fig:sup2}
\end{figure*}

\begin{figure*}
    \centering
    \includegraphics[width=15cm]{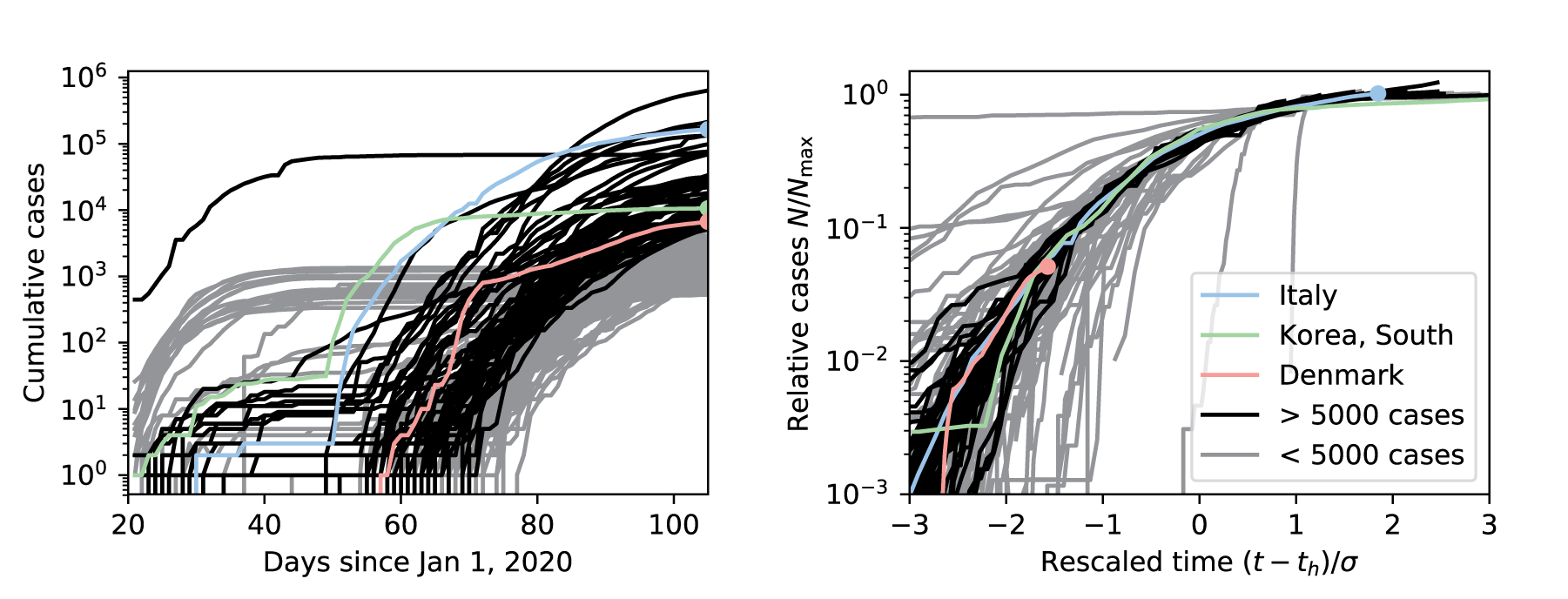}
    \caption{As Fig. \ref{fig:collapse}, but with cumulative case data instead of fatality data.}
    \label{fig:sup3}
\end{figure*}

\begin{figure*}
	\centering
	\includegraphics[width=15cm]{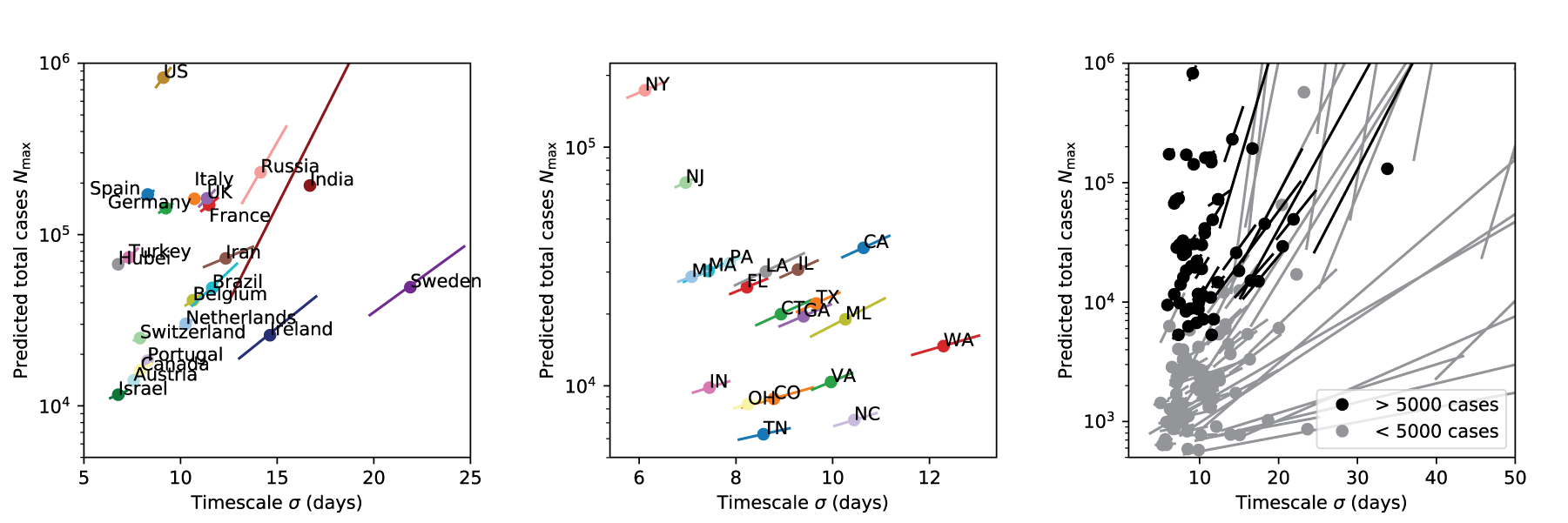}
	\caption{As Fig. \ref{fig:params}, but with cumulative case data instead of fatality data.}
	\label{fig:sup4}
\end{figure*}

\begin{figure*}
	\centering
	\includegraphics[width=15cm]{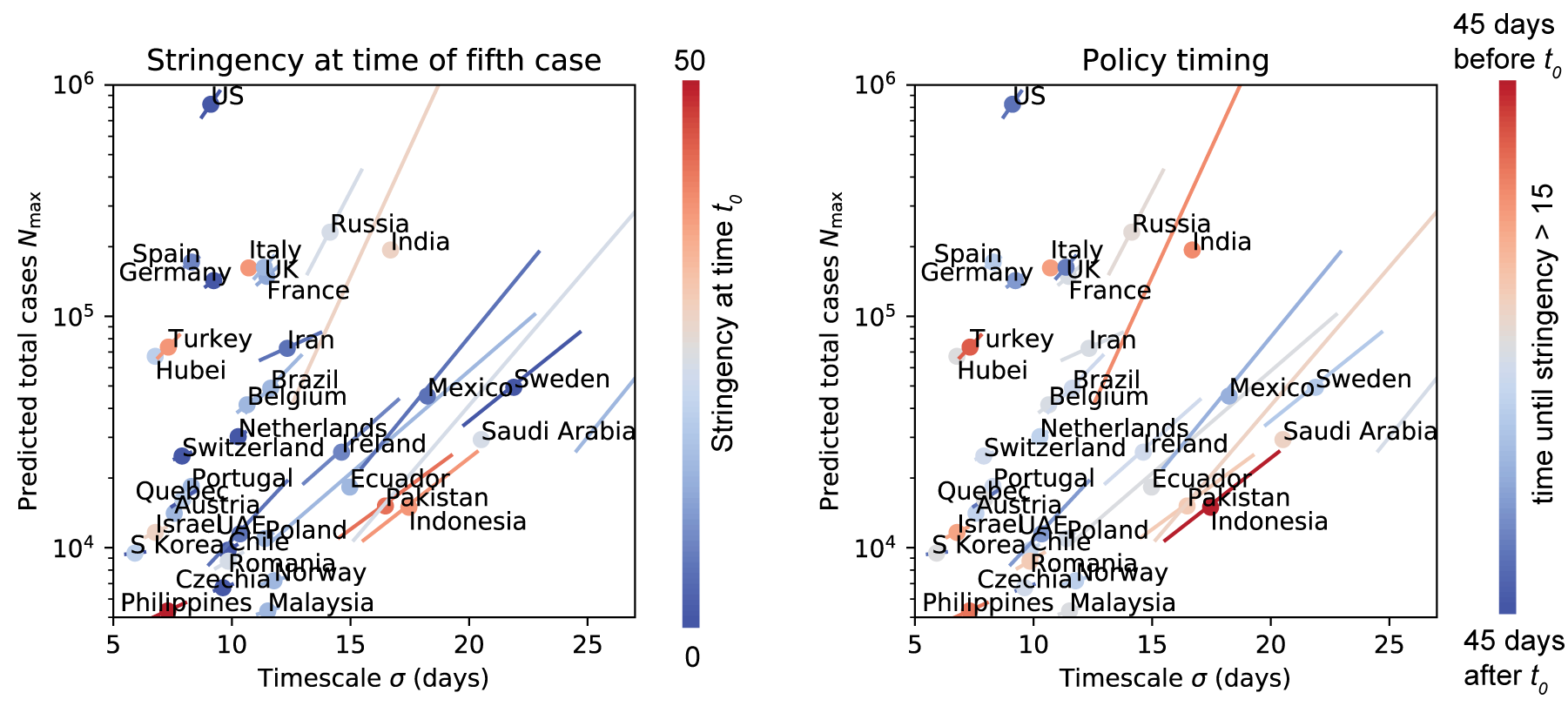}
	\caption{As Fig. \ref{fig:sd}, but with cumulative case data instead of fatality data. All countries with more than 5,000 cumulative cases as of April 15 are shown, with the exception of Japan and Peru. For these two countries, the case data was still compatible with continued exponential growth on April 15, and no estimate of $N_{\rm max}$ could be made. The error bar at the far right of the plot belongs to Denmark, whose best-fit value of $\sigma = 34$ lies outside the axis limits.}
	\label{fig:sup5}
\end{figure*}

\end{document}